\documentclass[aps,pra,reprint,superscriptaddress,floatfix]{revtex4-2}

\usepackage{graphicx}
\usepackage{amsmath,amssymb,amsfonts}
\usepackage{bm} % for bold math
\usepackage{float} % for controlling figure and table placement
\usepackage{color}
\usepackage{xcolor}
\usepackage{soul}
\usepackage{verbatim}
\setstcolor{red}
\usepackage{booktabs}
\usepackage{subcaption}
\usepackage[utf8]{inputenc}
\usepackage{multirow}
\usepackage{algorithm}
\usepackage{algorithmicx}
\usepackage{algpseudocode}
\usepackage{listings}
\usepackage{appendix}
\usepackage{caption}
\usepackage{ragged2e}
\usepackage{physics}
\makeatletter
\long\def\@makecaption#1#2{%
  \vskip\abovecaptionskip
  \noindent
  \begin{minipage}{\hsize}
    \raggedright
    #1. #2\par
  \end{minipage}
  \vskip\belowcaptionskip
}
\makeatother
\usepackage[
  colorlinks=true,
  linkcolor=blue,
  citecolor=blue,
  urlcolor=black,
  bookmarksopen=true
]{hyperref}
\usepackage[section]{placeins} % Prevent floats crossing sections

\usepackage{xcolor}
\usepackage[normalem]{ulem}
\usepackage{tikz}
\usetikzlibrary{decorations.markings}
\usetikzlibrary{calc}
\usetikzlibrary{arrows.meta, positioning}

\newcommand{\Eprint}[2]{}
\begin{document}
\preprint{APS/123-QED}

%\title{Scalability and Accuracy of Molecular Excited-State Calculations using Quantum Subspace Methods}
%\title{Practical Implementation of quantum subspace method q-sc-EOM with a discussion on resource requirements}
\title{Molecular Excited States using Quantum Subspace Methods: Accuracy, Resource Reduction, and Error-Mitigated Hardware Implementation of q-sc-EOM}
\author{Srivathsan Poyyapakkam Sundar} 
\affiliation{Department of Chemistry, University of North Dakota, ND 58202, USA}
\author{Prince Frederick Kwao} 
\affiliation{Department of Chemistry, University of North Dakota, ND 58202, USA}
\author{Alexey Galda} 
\affiliation{Moderna, Cambridge, MA 02141, USA}
\author{Ayush Asthana} 
\email{ayush.asthana@und.edu}
\affiliation{Department of Chemistry, University of North Dakota, ND 58202, USA}
\begin{abstract}

Problems in quantum chemical simulations, especially achieving accurate excited-state potential energy surfaces, are among the primary applications to achieve quantum utility. On near-term quantum hardware, variants of the variational quantum eigensolver (VQE) algorithms are the primary choice for chemistry simulation. In this study, a combination of leading ground and excited state quantum algorithms for general excited states, namely, ADAPT-VQE/LUCJ and q-sc-EOM, are utilized to calculate accurate excited state potential energy surfaces in challenging bond-breaking scenarios and compared with the classical scalable EOM-CCSD method. This work investigates avenues toward quantum utility in excited-state quantum chemistry using the q-sc-EOM approach. We assess its accuracy while mitigating major scaling bottlenecks through the Davidson algorithm and basis rotation grouping, reducing the measurement scaling from O(N$^{12}$) to O(N$^{5}$), and implementing the method on quantum hardware with various error mitigation strategies to reduce gate and measurement errors in excited states.  The hardware implementation of the q-sc-EOM algorithm, augmented by mitigation of M3 readout error and symmetry projection, produces reasonably accurate excited-state energies with gate noise identified as the predominant source of error. This paves the way for accurate and scalable, generally applicable quantum excited-state methods with potential for quantum utility while identifying critical problems that require advancements.
\end{abstract}
\maketitle
\section{Introduction}
Quantum-level simulation of eigenstates provides useful insights for the development of drugs, catalysts, materials and photochemical processes ~\cite{adamo2013calculations,ferre2016density} by accurately predicting molecular properties and reaction pathways. An example is in the development of photodynamic therapy for the treatment of various types of cancer, where computational chemistry methods could help in the development of improved photosensitizer molecules that could interact with light and produce the desired impact within the body ~\cite{zehr2025quantum}. Since predictive level excited-state calculations are challenging in highly correlated molecules, quantum computers may provide a meaningful benefit in the development and refinement of this class of problems. Also, excited-state quantum chemistry may be an ideal problem to explore quantum utility due to many unsolved challenges in the field that may be within the reach of early fault-tolerant quantum devices. 

Over the last several decades, many methods have been proposed that utilize a parallelized classical computing infrastructure to make accurate predictions of ground and excited state properties of molecules. The primary approaches used currently are based on Density Functional Theory (DFT) ~\cite{doi:10.1021/acs.jpca.1c08310,doi:10.1021/acs.jpclett.1c00744} and wavefunction theory, especially coupled-cluster singles and doubles with perturbative triples (CCSD(T)) ~\cite{haakansson2020bromide,watson2016anion,miyagawa2019domain,kozma2020new}. Although widely useful, these methods generally fail when the ground state becomes multi-reference, which often arises in bond breaking, systems with multiple transition elements (or heavy elements), and the potential energy surface of electronically excited states.
Progress has been made for systems with strong correlation, where CASPT2 ~\cite{tran2021geometric,gil2020synthesis,heitz1995caspt2,yamanaka1994caspt2} and DMRG ~\cite{sharma2019density,baiardi2020density,alvarez2009density} are leading methods, but they are limited in their reach.
For the calculation of excited-state energies and properties, numerous classical methods have been developed in the past~\cite{szalay2012multiconfiguration,stanton1993equation,sneskov2012excited,mayhall2014quasidegenerate,nooijen1997new,mazziotti2003extraction}. 
The most popular scalable methods are Time-dependent DFT (TD-DFT)~\cite{bauernschmitt1996treatment,stratmann1998efficient} and equation of motion-coupled cluster (EOM-CCSD)~\cite{stanton1993equation}. 
These methods provide meaningful insight in excited-state processes but face challenges at conical intersections~\cite{rossi2025generalized,10.1063/5.0041822}, doubly excited states~\cite{damour2024state,ravi2023excited}, or when the ground state has a strong multi-reference character~\cite{schmidt1998construction,kohn2013state}. 

Quantum computers can play a vital role in excited-state quantum chemistry by providing an alternate route to solving the exponentially scaling exact molecular electronic structure problem, and therefore, providing accurate predictive computations for molecular properties.  
The primary workhorse for near-term quantum computers is variational quantum eigensolver (VQE)~\cite{PhysRevA.95.020501, hempel2018quantum,kandala2017hardware,o2016scalable}. 
To reach excited states using near term quantum computers, several methods have been proposed ~\cite{motta2020determining,parrish2019quantum,sugisaki2022adiabatic,nakanishi2019subspace,o2019quantum,bauman2020toward,santagati2018witnessing,russo2021evaluating,chan2021molecular,higgott2019variational}, where diagonalisation-based approaches have gained significant importance.
 This is because, compared to other approaches, they are usually reasonable in resource requirements, provide accurate excited states and are unaffected by degeneracies and near-degeneracies. 
 The popular diagonalization-based approaches are  Quantum Subspace Expansion (QSE)~\cite{colless2018computation,mcclean2020decoding,takeshita2020increasing}, quantum equation-of-motion (qEOM)~\cite{ollitrault2020quantum} and  Quantum self-consistent equation of motion (q-sc-EOM)~\cite{asthana2023quantum}.
Among the several options, the q-sc-EOM method (and related linear response formalism~\cite{kumar2023quantum,ziems2025understanding}) is shown to be rigorous theoretically as it has the correct scaling property (size-extensivity) in excitation energies, ensures correct orthogonality of states by satisfying the killer condition, and is significantly more robust to measurement errors due to its orthogonal basis.

Q-sc-EOM requires preparing an accurate ground-state through a VQE subroutine. Many ansätze have been proposed for the state preparation task within VQE. 
Hardware-efficient ansatz (HEA) are efficient but face difficulty in avoiding barren plateaus~\cite{zhuang2024hardware,wang2021noise,mcclean2018barren}. For molecular problems, the Unitary Coupled Cluster Singles and Doubles (UCCSD)~\cite{peruzzo2014variational,barkoutsos2018quantum,anand2022quantum,yung2014transistor} ansatz is promising as it is built using the well-studied unitary coupled cluster theory, but is often unable to reach accurate energies in strongly correlated problems~\cite{anand2022quantum,PhysRevResearch.6.013254,wu2024multi,fedorov2022vqe}. 
A promising approach for challenging problems is the Adaptive Derivative Assembled Pseudo Trotter ansatz Variational Quantum Eigensolver (ADAPT-VQE) ansatz that was developed by Grimsley et al.~\cite{grimsley2019adaptvqe}. 
ADAPT-VQE started the family of dynamic ansätze and it builds the ansatz one at a time by adding an operator with the highest gradient at each iteration~\cite{anand2022quantum,mullinax2024classical,PRXQuantum.2.020310}, but it has limitations for application to the currently available hardware due to its high measurement requirements. The available ansatz suitable for hardware application is Local Unitary Cluster Jastrow (LUCJ) ansatz, which is usually used with precomputed parameters from coupled cluster theory.
A recent alternative approach to VQE that is another promising direction for near-term hardware is the recent sample-based quantum diagonalization (SQD) technique ~\cite{shajan2024towards,barison2025quantum,shivpuje2025sample} that has demonstrated that it can reach accurate ground and excited-state energies on quantum hardware.

In the first phase of this work, we calculated the challenging ground and excited potential energy surfaces with the ADAPT-VQE+q-sc-EOM strategy using quantum simulators with exact and sampling noise inclusion to assess the accuracy of the method on perfect quantum hardware. The key limitation of this strategy is the high shot count scaling, O(N$^{12}$), with the increasing number of orbitals (N). In the second phase of this work, we proposed the use of Davidson algorithm developed by Kim et al. ~\cite{kim2023two} and the basis rotation grouping (BRG)~\cite{Huggins2021, Gonthier2022, kojima2025orbital} that uses low-rank factorization of the two electron tensor (g$_{\text{pqrs}}$) introduced by Huggins et al. ~\cite{huggins2021efficient}. This strategy can bring down the effective shot count scaling cost to O(N$^5$), making the excited-state calculations much more scalable. Finally, in the third phase, we implemented the q-sc-EOM algorithm with a ground-state prepared by an LUCJ ansatz on quantum hardware, tested several error mitigation strategies and reached excited-state energies with reasonable accuracy; however, progress in hardware and error mitigation strategies will be necessary to reach excited-state properties useful for spectroscopy and other applications.

The remainder of the manuscript is organized as follows. In Sec .~\ref {Theory}, we provide the theoretical background for the ADAPT-VQE, LUCJ and the q-sc-EOM algorithms. In Results and Discussion Sec \ref{rd1}, we compare the excited-state energies with sampling noise against the classical algorithm EOM-CCSD for $\mathrm{NH_3}$ and $\mathrm{H_2O}$ molecules. In Sec \ref{rd2}, we discuss the resource requirements and provide strategies to reduce the overall shot count scaling,  and in Sec \ref{rd3}, we conclude by depicting hardware results upon implementing several error mitigation techniques. Finally, in Sec.\ref{Conclusion}, we discuss conclusions of the study and the broader implications of the excited-state algorithm q-sc-EOM using quantum hardware.

\section{Theory}\label{Theory}
ADAPT-VQE, LUCJ and q-sc-EOM algorithms are used in this work for ground-state and excited-state calculations. 
The codes were developed in-house using Pennylane software ~\cite{bergholm2018pennylane} and PYSCF software, and made available in our Git repository linked in the data availability section.
The operator pool was constructed using all UCCSD-style fermionic single and double excitations. 
The active space in each molecular case is noted in the following sections.  
The Hartree-Fock state is the initial state in all the computations in this study.

\subsection{ADAPT-VQE}
The ADAPT-VQE algorithm produces a dynamic ansatz specifically to that system. It does that by measuring the gradient of each operator in a pool of predefined operators and adding the operator with the highest gradient to the circuit at each iteration with a new variational parameter (${\theta}$). It then carries out an optimzation routine to optimize the parameters associated with each added operator in the circuit. Gradients are computed by calculating the commutator of the Hamiltonian with respect to each of the operators ($\hat{O}$) present in the pool, as
\begin{align}
    \frac{\partial E^{(m)}}{\partial \theta}= \bra{\Psi_{\text{m}}}[\hat{H},\hat{O}]\ket{\Psi_{\text{m}}}.
\end{align}
The new wavefunction is given by
\begin{align}
    \ket{\Psi_{\text{(m+1)}}} = e^{-\theta_{m+1} \hat{O}_{m+1}}\ket{\Psi_{\text{(m)}}}.
\end{align}
 The final output state is obtained once operators have been added until the norm of the gradient vector drops below a specified threshold.
%This ADAPT-VQE algorithm outperforms UCCSD with lower parameters and circuit depth and also achieves higher accuracy for strongly correlated systems. 
\subsection{LUCJ ansatz}
The Local Unitary Cluster Jastrow (LUCJ) ansatz is a hardware-aware, chemically motivated alternative to standard qUCC-style circuits for ground-state preparation in strongly correlated regimes~\cite{motta2023lucj,motta2024quantum}. The central idea is to retain the physical structure of orbital rotations and correlation-inducing two-body interactions, while constraining entangling operations to local qubit connectivity so that circuit depth and SWAP overhead are reduced on near-term architectures.

In a layered form, the wavefunction can be expressed as
\begin{align}
    \ket{\Psi_{\mathrm{LUCJ}}} = \prod_{\mu=1}^{L} e^{\hat{K}_{\mu}} e^{\hat{J}_{\mu}}\ket{\Phi_0},
\end{align}
where \(\ket{\Phi_0}\) is typically the Hartree--Fock determinant, \(\hat{K}_{\mu}\) is an orbital-rotation generator (number-conserving one-body anti-Hermitian operator), and \(\hat{J}_{\mu}\) is a unitary Jastrow correlator written in terms of number operators. In practical implementations, locality constraints are imposed on \(\hat{J}_{\mu}\), yielding the ``local'' UCJ form that maps efficiently to sparse device coupling graphs.

Compared to UCCSD circuits, LUCJ provides a more favourable compromise between expressibility and hardware cost: it captures strong correlation effects with fewer long-range entangling operations and can be systematically improved by increasing the number of layers \(L\)~\cite{motta2023lucj}. Recent optimization studies using stochastic reconfiguration/linear-method updates also indicate improved robustness for correlated electronic-state optimization, which is useful for obtaining high-quality reference states for subspace excited-state methods such as q-sc-EOM~\cite{motta2024quantum}.

\subsection{q-sc-EOM}
For excited-states, the quantum self-consistent equation of motion (q-sc-EOM) method calculates the excited-state energies using the equation of motion formalism. Q-sc-EOM method is a formulation that uses self-consistent operators in the q-EOM framework developed by Pauline et al.~\cite{ollitrault2020quantum}. The governing equation of the q-sc-EOM method is given by
\begin{align}
    \mathbf{M}C=CE,
\end{align}
where the matrix elements of $\mathbf{M}$ are 
\begin{align}
    \mathbf{M}_{I,J}=\bra{{\text{HF}}}\hat{G}_I^{\dagger}U^{\dagger}(\theta)\hat{H}U(\theta)\hat{G}_J\ket{{\text{HF}}} - \delta_{I,J}E_{\text{HF}}
\end{align}
where $\hat{G}_I$ and $\hat{G}_J$ are the excitation manifolds that can be developed by including possible single and double excitations. 
\begin{align}
    \begin{split}
    \hat{G}_{j}^{a} &= \hat{a}^{\dagger}_a\hat{a}_j,\\
    \hat{G}_{ji}^{ab} &= \hat{a}^{\dagger}_a\hat{a}^{\dagger}_b\hat{a}_i\hat{a}_j,
    \end{split}
\end{align}
Here, $U(\theta)$ is a parameterized quantum circuit, obtained from the ground-state energy calculations. 
\begin{align}
    \ket{\Psi_{\text{VQE}}}=U(\theta)\ket{{\text{HF}}}
\end{align}
Finally, the eigenvalue equation is solved by diagonalizing the M matrix to compute the excited-state energies of the molecule.

\section{Results and Discussion} \label{RandD}
\subsection{Accuracy analysis compared with scalable classical simulations}\label{rd1}
Quantum utility will require a quantum algorithm to outperform the best classical algorithms. Here we perform an analysis with a strategy of ADAPT-VQE for ground state and q-sc-EOM for the excited state, with sampling errors included in the simulations to assess the possibility of reaching quantum utility in excited-state quantum chemistry problems. This analysis assumes perfect quantum hardware, but even perfect hardware will have sampling errors because these originate at the fundamental level of the quantum nature of the hardware. Below are examples of two important ``two bond-breaking'' cases.

\subsubsection{\texorpdfstring{$NH_3$}{NH3}: two bonds breaking case}
\begin{figure*}[t]
    \centering
    \begin{subfigure}[b]{0.45\textwidth}
    \includegraphics[width=\linewidth]{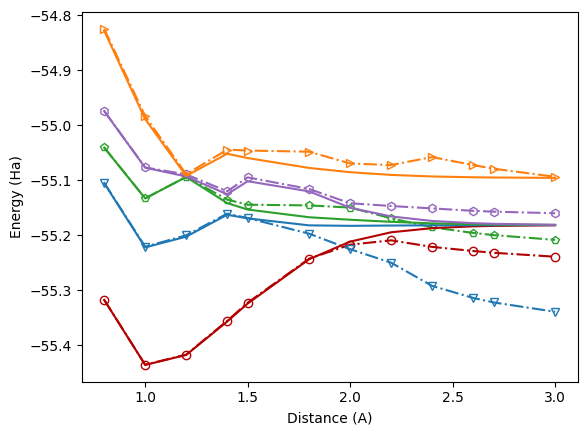}
    \caption{EOM-CCSD vs FCI}
    \label{a_cvsq}
    \end{subfigure}
    \begin{subfigure}[b]{0.45\textwidth}
    \includegraphics[width=\linewidth]{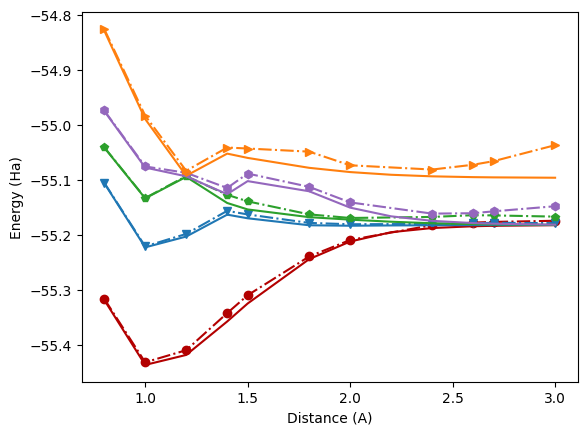}
    \caption{ADAPT-VQE + q-sc-EOM vs FCI}
    \label{b_cvsq}
\end{subfigure}
\caption{Potential energy surface of ground and excited states as a function of bond distance in $\mathrm{NH}_3$ with two N-H bond breaking calculated using (a) classical algorithm EOM-CCSD and (b) quantum algorithm ADAPT-VQE + q-sc-EOM, represented using a combination of dashed lines and symbols. FCI energy values for all the states are represented by solid lines.}
\end{figure*}

The potential energy surface of $\mathrm{NH}_3$ with two bonds dissociating simultaneously, with the ground-state energy and its wave function $\mathbf{\ket{\psi_{\mathrm{VQE}}}}$  generated using ADAPT-VQE and excited states using q-sc-EOM, represented in Fig.\ref{b_cvsq}. The results of FCI ground and excited states are represented using solid lines. For $\mathrm{NH}_3$, we selected an active space considering six electrons in six spatial orbitals. This becomes a 12-qubit system with 117 possible single and double excitations. Out of which, 25 excitations were chosen based on gradient calculations and added to the Hartree Fock state and their parameters were optimized in the ADAPT-VQE algorithm, generating the ground-state energy and the unitary matrix that is necessary for the excited-states calculation. The results of q-sc-EOM overlap with the FCI energies.  Since ADAPT-VQE captures most of the correlation energy even in strongly correlated regimes (refer to Fig.\ref{1}), led to more accurate results in the excited states. 

To benchmark the performance of the quantum algorithm, a comparison with a scalable classical algorithm like EOM-CCSD is essential. The ground and excited states for $\mathrm{NH_3}$ using the classical method are represented in Fig.\ref{a_cvsq}. CCSD ground-state energies align with FCI data till 2$A^o$. Once the system enters the regime of strong correlation, multi-reference character becomes significant, thereby CCSD fails to accurately reproduce the ground-state properties, in agreement with the results shown in Fig.\ref{1}. Inadequate description of the ground state led to inappropriate results in the excited states in EOM-CCSD. 

\begin{figure}[htbp]
    \centering
    \includegraphics[width=\linewidth]{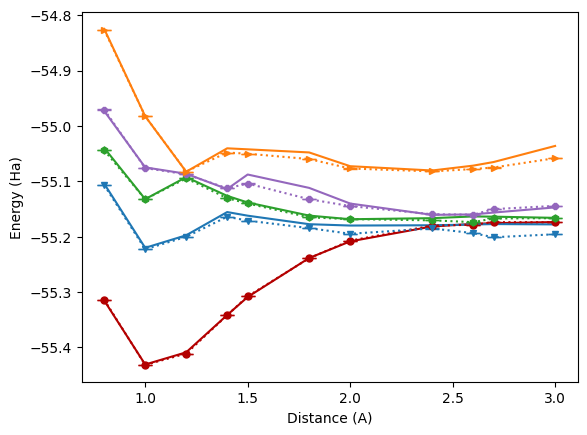}
    \caption{Potential energy surface of ground and excited states as a function of bond distance in $\mathrm{NH}_3$ with two N-H bond breaking calculated using quantum algorithm ADAPT-VQE + q-sc-EOM with 10000 shots, presented using a combination of symbols and dotted lines. Exact energy values for all the states are represented by solid lines.}
    \label{4}
\end{figure}

Statistical sampling effects are an inherent component of a perfect quantum computer. In this study for excited states, we introduce the effect of noise by adding statistical sampling effects (shot noise) to evaluate the performance of q-sc-EOM algorithm. In the ground state algorithm, the operator selection process and its parameter optimization were performed in an exact scenario (infinite shots). The noise is introduced only in the measurements of energy. The other sources of errors (bit, phase flip, and gate noise) will be discussed in detail in the hardware results section. Fig.\ref{4} represents the performance of the q-sc-EOM algorithm in the presence of shot noise. Fig.\ref{4} shows the energy averaged over multiple calculations for 10k shots per matrix element.  Upon comparison of the quantum algorithm (ADAPT-VQE + q-sc-EOM with shots) to the classical system (EOM-CCSD, refer Fig.\ref{a_cvsq}), we find that doubly excited states are accurately described using quantum algorithm even in the presence of shot noise in comparison to EOM-CCSD. 

From the results (Fig.\ref{4}), the inference is that quantum algorithm (all VQE-based systems), i.e. Adapt-VQE, inherently holds multi-reference character and could accurately capture ground-state energies, thereby leading to more accurate excited state energies than the competitive classical method. 
\subsubsection{\texorpdfstring{H$_2$O}{H2O}: two bond breaking case}
\begin{figure*}[t]
    \centering
    \begin{subfigure}[b]{0.48\textwidth}
    \includegraphics[width=\linewidth]{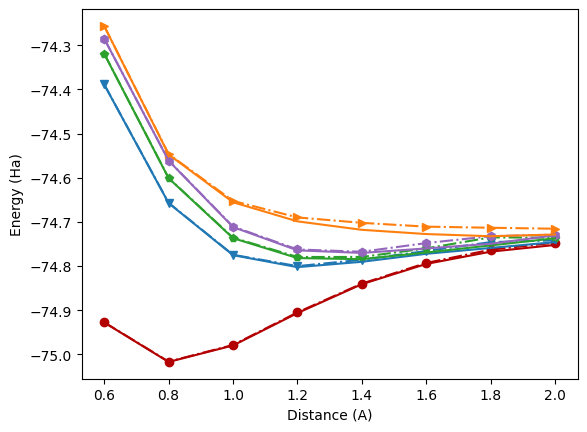}
    \caption{ADAPT-VQE + q-sc-EOM (exact) vs FCI}
    \label{5}
    \end{subfigure}
    \hfill
    \begin{subfigure}[b]{0.48\textwidth}
    \includegraphics[width=\linewidth]{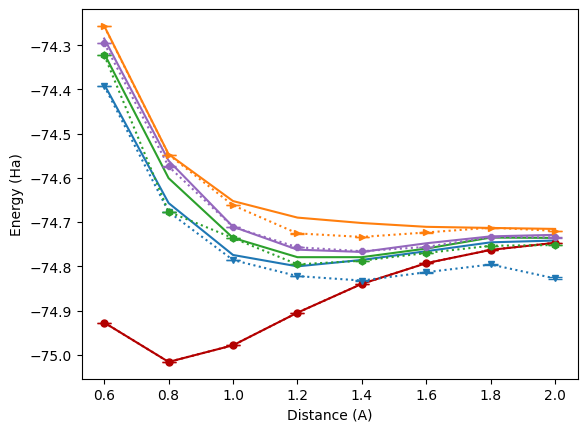}
    \caption{ADAPT-VQE + q-sc-EOM (10,000 shots) vs exact}
    \label{6}
    \end{subfigure}
    \caption{Potential energy surface of $\mathrm{H}_2\mathrm{O}$ representing ground and excited states as a function of bond distance: (a) quantum algorithm ADAPT-VQE + q-sc-EOM in an exact scenario compared against FCI, and (b) ADAPT-VQE + q-sc-EOM with 10,000 shots compared against exact values.}
\end{figure*}

Quantum algorithm (Adapt + q-sc-EOM) provided accurate description of excited states for $\mathrm{NH}_3$ with 12 qubits. We further decides to test for a larger system, $\mathrm{H}_2\mathrm{O}$. In this study, the full space was considered as active with 10 electrons in 7 spatial orbitals. The potential energy surface of two H bonds dissociating simultaneously in $\mathrm{H}_2\mathrm{O}$ is modeled in an exact scenario and is represented in Fig.\ref{5}. The energies obtained using the quantum algorithm are accurate upon comparison to FCI energies. Then, statistical sampling effects are introduced into the q-sc-EOM with 10k shots per matrix element and are compared with exact simulation of the quantum algorithm, represented in Fig.\ref{6}. The results show that there is a necessity to include more shot counts to accurately model the system. 

Based on the results of $\mathrm{NH}_3$ and $\mathrm{H}_2\mathrm{O}$, we can conclude that the quantum algorithm (Adapt + q-sc-EOM) produces accurate excited-state potential energy surfaces, sometimes even more accurate than the classical scalable EOM-CCSD method, providing initial hope of quantum utility in excited state quantum chemistry using subspace diagonalization based approaches provided high quality hardware is available. 
Study on other sources of noise and hardware implementation are included in the later sections and analyzed.

\subsection{Resource reduction strategies}\label{rd2}
%Oneliner: Multiple strategies can be implemented such as Davidson and Basis rotation grouping to reduce the scaling of q-sc-EOM calculation by orders of magnitude from $\mathrm{O(N^{12})}$ to $\mathrm{O(N^{5})}$. 
%q-sc-EOM main scaling 

We initiate by exploring the resource requirements for the q-sc-EOM algorithm. The excited state energies are computed by the q-sc-EOM algorithm utilizing both quantum and classical systems. 
%The total number of matrix elements that needs to be measured on a quantum device depends on the level of the excitations used. 
The computational bottleneck lies in the number of matrix elements that need to be evaluated on the quantum hardware. If the excitations are constrained to singles and doubles excitation, measurement scales in the order of $\mathrm{O(N^4 m_{occ}^{4} m_{virt}^4)}$ where $\mathrm{N},\mathrm{m_{occ}, m_{virt}}$ refer to number of qubits, and the number of occupied and virtual spin orbitals, respectively. Hence, the overall scaling of the brute force implementation of q-sc-EOM algorithm is $\mathrm{O(N^{12})}$. 

Two strategies can be implemented, either by reducing the shot scaling for the Hamiltonian terms or by reducing the resource for the number of occupied and virtual orbitals. 
\subsubsection{Davidson algorithm}
To reduce the measurement cost, Kim et al.~\cite{kim2023two} adapted the classical Davidson diagonalization algorithm for the quantum computations.  The implementation of the Davidson algorithm leads to a significant reduction in the scaling from $\mathrm{O(N^4 m_{occ}^{4} m_{virt}^4)}$ to $\mathrm{O(N^4 m_{occ}^{2} m_{virt}^2)}$. Rather than the evaluation of all matrix elements, this Davidson procedure initiates with a span of guess vectors and iteratively expands towards the target specific subspace based on the feedback from the Davidson procedure, by measuring the residual vectors.

There is a projection step involved in this procedure to project the residual vectors to single and double excited determinants, which leads to the measurement of overlaps between the excited Slater determinants and the residual vectors, leading to the above scaling. In terms of system size using the more generic $N$ representing a general number of orbitals, this corresponds to a reduction in the overall algorithmic scaling from the order of $\mathrm{N^{12}}$ to approximately $\mathrm{N^{8}}$, representing a significant improvement in the feasibility of q-sc-EOM approach for larger molecular systems. 
This shot count scaling benefit arose at the cost of an increase in circuit depth that scales as O($N^4$) in the number of multi-control NOT gates. Further, the implementation of multi-control NOT gates requires O(N) or O(N$^2$) (depending on whether ancilla qubits are used or not) number of CNOT gates, making the overhead of CNOT gates as O(N$^5$) or O(N$^6$). It should be noted that this procedure is exact and has no impact on the accuracy of the calculations, and one can select the number of states of interest and achieve the standard q-sc-EOM accuracy.

\begin{figure}[htbp]
    \centering
    \includegraphics[width=0.95\linewidth]{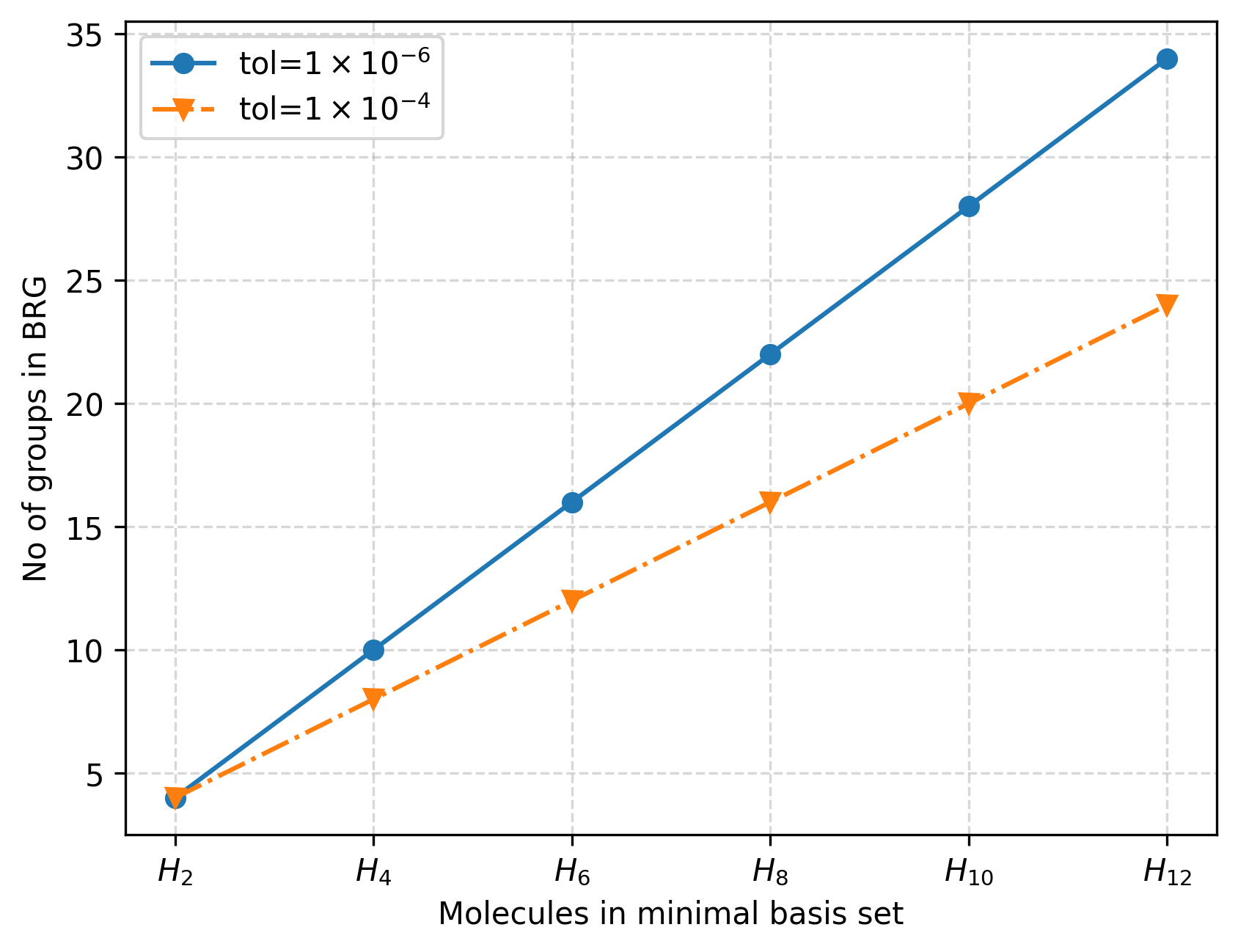}
    \caption{Number of BRG groups as a function of molecular hydrogen chain size ($H_2$, $H_4$, $H_6$, $H_8$, $H_{10}$, $H_{12}$)with multiple tolerance values in a minimal basis set.}
    \label{BRG_Groups_Hchains}
\end{figure}

\begin{figure*}[t]
    \centering
    \includegraphics[width=0.45\textwidth]{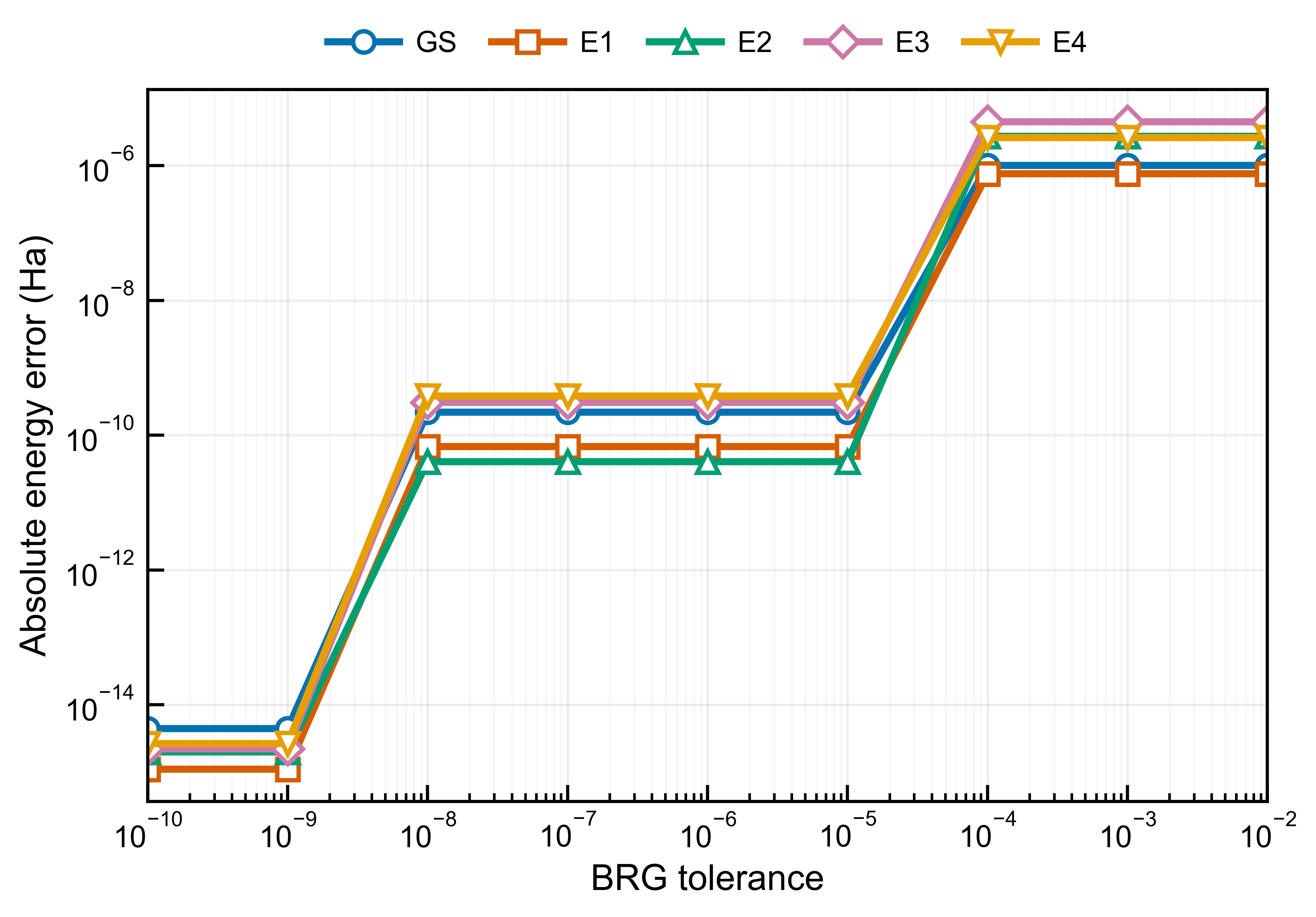}
    \hfill
    \includegraphics[width=0.45\textwidth]{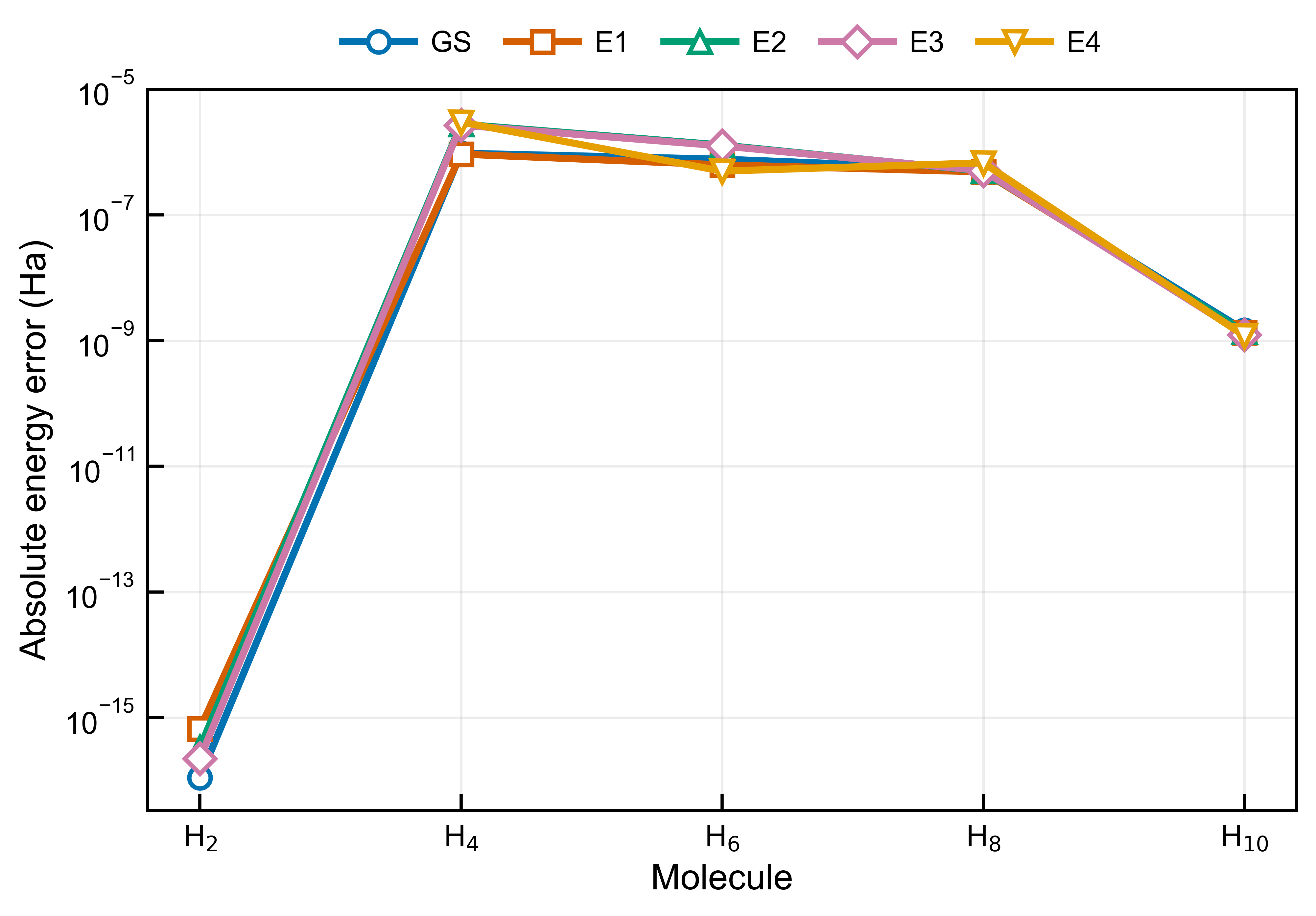}
    \caption{Left: absolute q-sc-EOM excitation-energy errors for linear H$_4$ in STO-3G at 1.5~\AA{} as a function of BRG tolerance. Right: absolute q-sc-EOM energy errors for linear H chains from H$_2$ to H$_{10}$ in STO-3G at 1.5~\AA{} using fixed BRG tolerance $10^{-4}$.}
    \label{fig:qsceom-brg-h4-h10}
\end{figure*}

\subsubsection{Basis rotation grouping}

Multiple techniques are proposed in the literature to carry out effective reduction of the number of Pauli strings in the Hamiltonian that needs to be measured in the calculation of expectation values of Hamiltonian operators, including QubitWise Commutativity (QWC)~\cite{PhysRevA.110.022606, 10.1063/1.5141458} and BRG. The BRG technique exploits the sparsity of the Hamiltonian. The molecular Hamiltonian is a combination of one $\mathrm{h_{pq}}$ and two-body integrals $\mathrm{g_{pqrs}}$, 
\begin{align}
\begin{split}
\hat H
&=
\sum_{\alpha\in\{\uparrow,\downarrow\}}\sum_{ij}
h_{ij}\, c^\dagger_{i\alpha} c_{j\alpha}
\\&+
\frac12
\sum_{\alpha,\beta\in\{\uparrow,\downarrow\}}\sum_{ijkl}
g_{ijkl}\,
c^\dagger_{i\alpha} c^\dagger_{k\beta} c_{l\beta} c_{j\alpha},
\label{eq:H_standard}
\end{split}
\end{align}
where $g_{ijkl}$ implies a two-electron integral in the chemist notation. The technique developed by Huggins et al.~\cite{Huggins2021} maps the two-body tensor $\mathrm{g_{pqrs}}$ to low rank tensors, reducing the scaling of measurement of expectation value of Hamiltonian from $\mathrm{O(N^4)}$ to $\mathrm{O(N)}$, a cubic improvement over the brute force method. The two-electron tensor $g$ is converted to a low-rank tensor indexed by the composite indices $ij$ and $kl$,  and eigen-decomposed up to rank $R$ to give
\begin{equation}
    g_{ijkl}
=
\sum_{r=1}^{R}
L^{(r)}_{ij}\, L^{(r)}_{kl},
\end{equation}
where $L$ denotes the matrix of eigenvectors of the matrix $g$. The new Hamiltonian can be written as
\begin{align}
\begin{split}
\hat H&
=
\sum_{\alpha\in\{\uparrow,\downarrow\}}\sum_{ij}
\tilde h_{ij}\, c^\dagger_{i\alpha} c_{j\alpha}
\\&+
\frac12
\sum_{r=1}^{R}
\left(
\sum_{\alpha\in\{\uparrow,\downarrow\}}\sum_{ij}
L^{(r)}_{ij}\, c^\dagger_{i\alpha} c_{j\alpha}
\right)^2.
\end{split}
\end{align}
The orbital basis can be rotated such that each $h$ and $L^{(r)}$ matrix is diagonal. 
The Hamiltonian can then be written as
\begin{equation}
    \hat H
=
U_0
\left(\sum_i d_i\, n_i\right)
U_0^\dagger
+
\frac12
\sum_{r=1}^{R}
U_r
\left(
\sum_i \lambda_i^{(r)} n_i
\right)^2
U_r^\dagger,
\end{equation}
where the diagonalization of the $h$ and $L^{(r)}$ matrices, leads to the coefficients $d_i$.

This formulation in the qubit form is explored here for the use of the BRG technique in the q-sc-EOM algorithm. In Fig. \ref{BRG_Groups_Hchains}, it can be seen that, as seen by other studies, the number of groups scales linearly with the varying molecular size for both of the chosen tolerance values. A stricter tolerance ($\mathrm{1\times10^{-6}}$) retains more features of the electron correlation, leading to more groups, while the looser tolerance trims smaller features, thereby producing fewer groups. For $\mathrm{H_{12}}$, this difference sums to approximately 10 more groups under stricter tolerance, highlighting the trade-off between measurement cost and accuracy. Further, in fig.~\ref{fig:qsceom-brg-h4-h10} we study the accuracy of ground and excited states of linear H chains starting from $\mathrm{H_2}$  to $\mathrm{H_{10}}$ using a fixed tolerance value of $\mathrm{10^{-4}}$ showing that errors are well below $\mathrm{10^{-6}}$ Ha across the ground and excited states, stating that BRG preserves the accuracy of both the ground and excited states using q-sc-EOM.

\begin{figure*}[t]
    \centering
    \begin{subfigure}[b]{0.45\textwidth}
    \includegraphics[width=\linewidth]{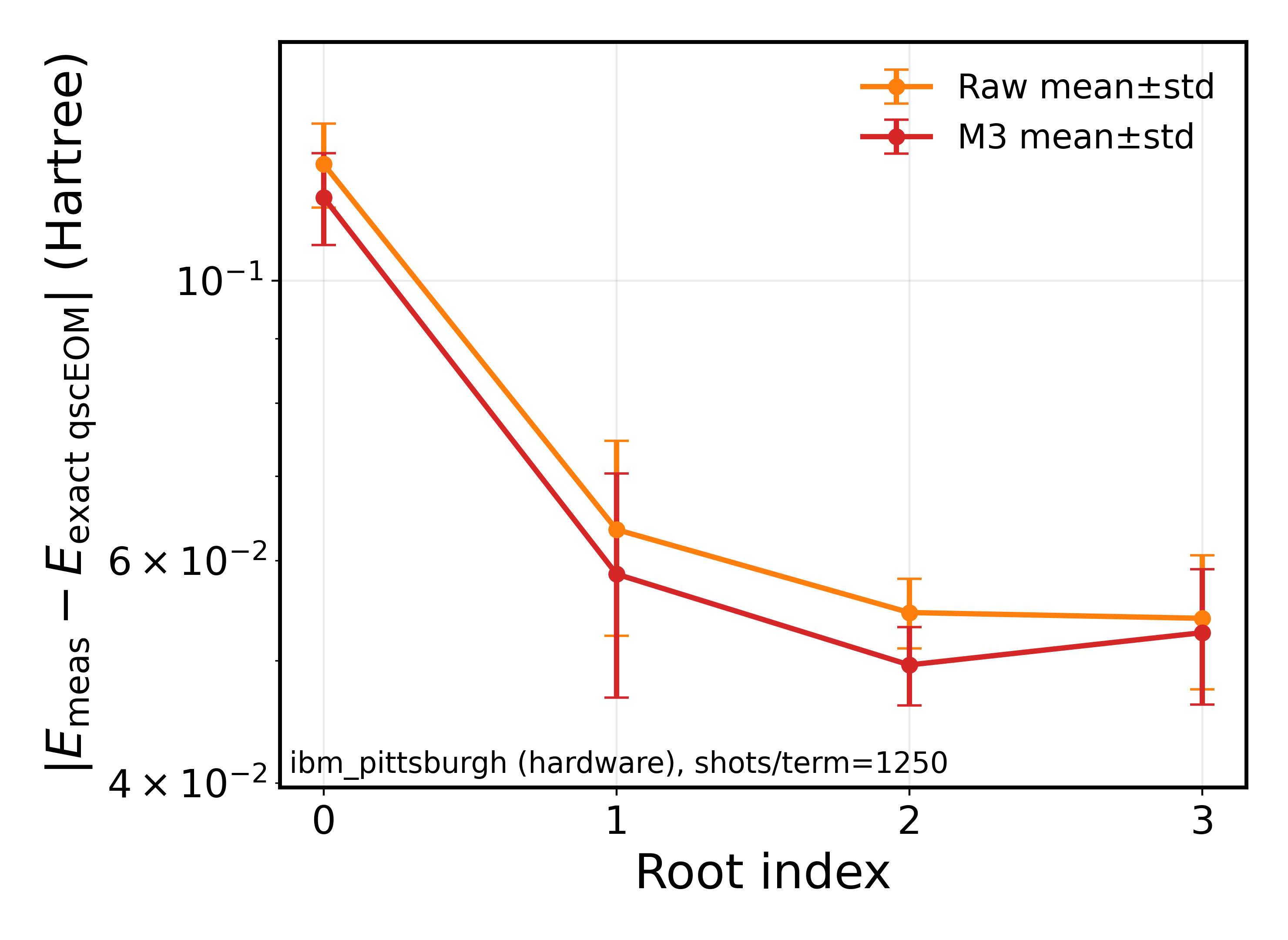}
    \caption{Error vs root index}
        \end{subfigure}
    \begin{subfigure}[b]{0.45\textwidth}
    \includegraphics[width=\linewidth]{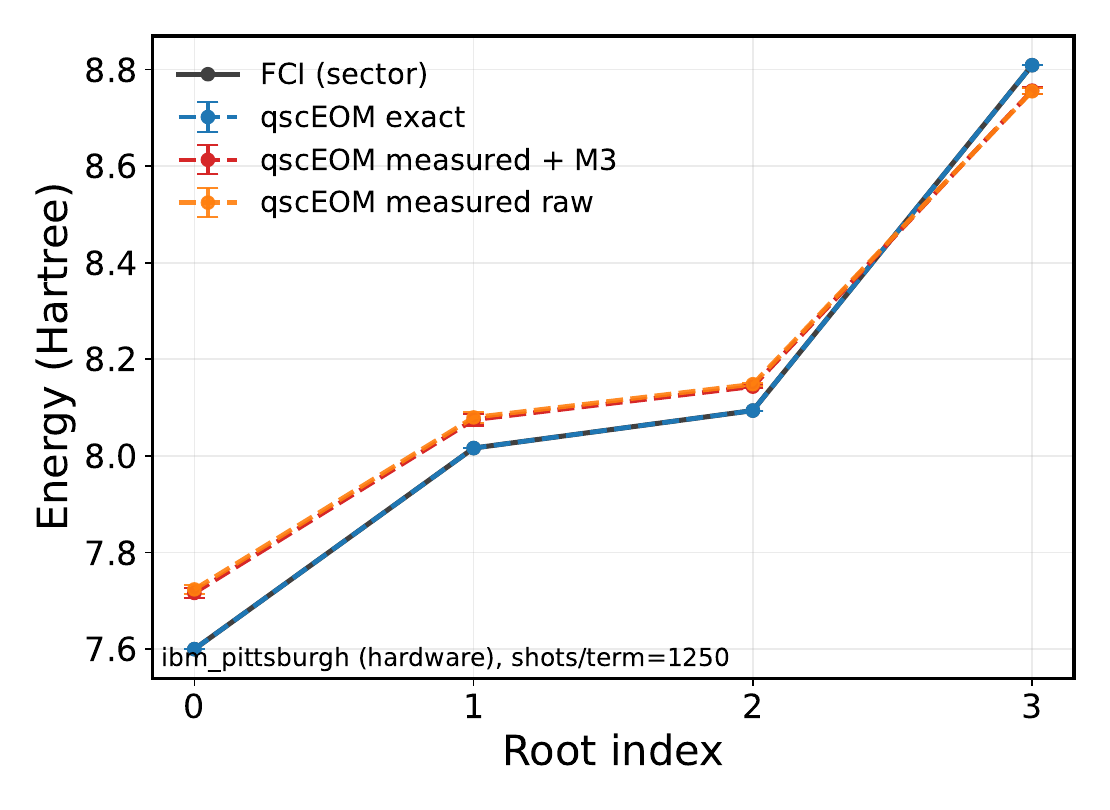}
    \caption{Energy vs root index}
    \end{subfigure}
\caption{Eigen energies of ground and excited states of H$_2$O (active space of (2e,2o), STO-3G basis with a bond length 0.94$\text{\AA}$) with a total shot count of 100,000 on IBM Pittsburgh quantum hardware using 5 runs. Only Pauli grouping and M3 error mitigation are used in the calculation, which is represented by the red line. The calculations are able to reach reliable accuracies of approximately 50 mHa for excited state energies using q-sc-EOM.}
\label{fig:h2o_hw_roots}
\end{figure*}
\begin{figure*}[htbp]
    \centering
    \begin{subfigure}[b]{0.45\textwidth}
    \includegraphics[width=\linewidth]{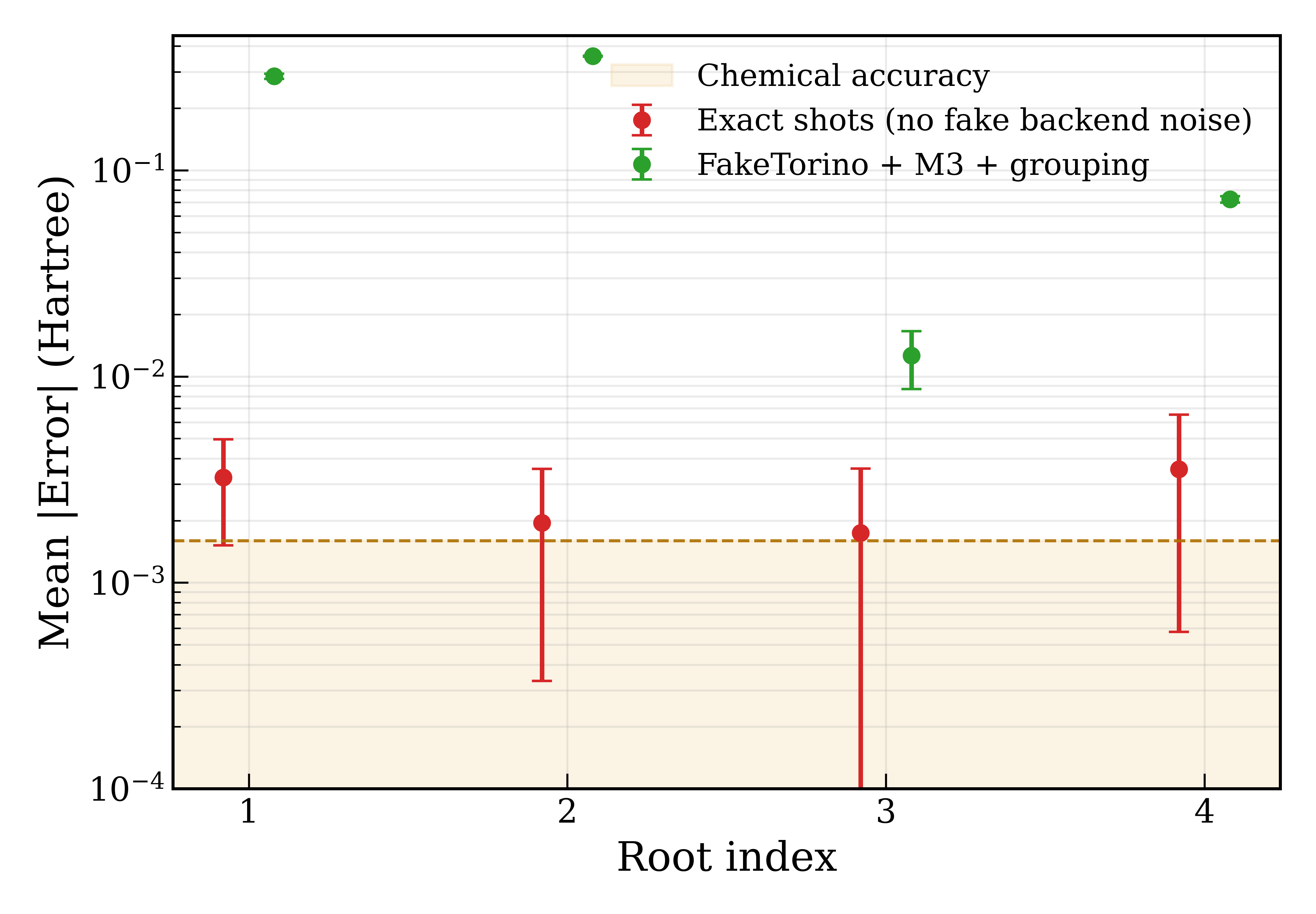}
    \caption{Fake Torino device }
        \end{subfigure}
    \begin{subfigure}[b]{0.45\textwidth}
    \includegraphics[width=\linewidth]{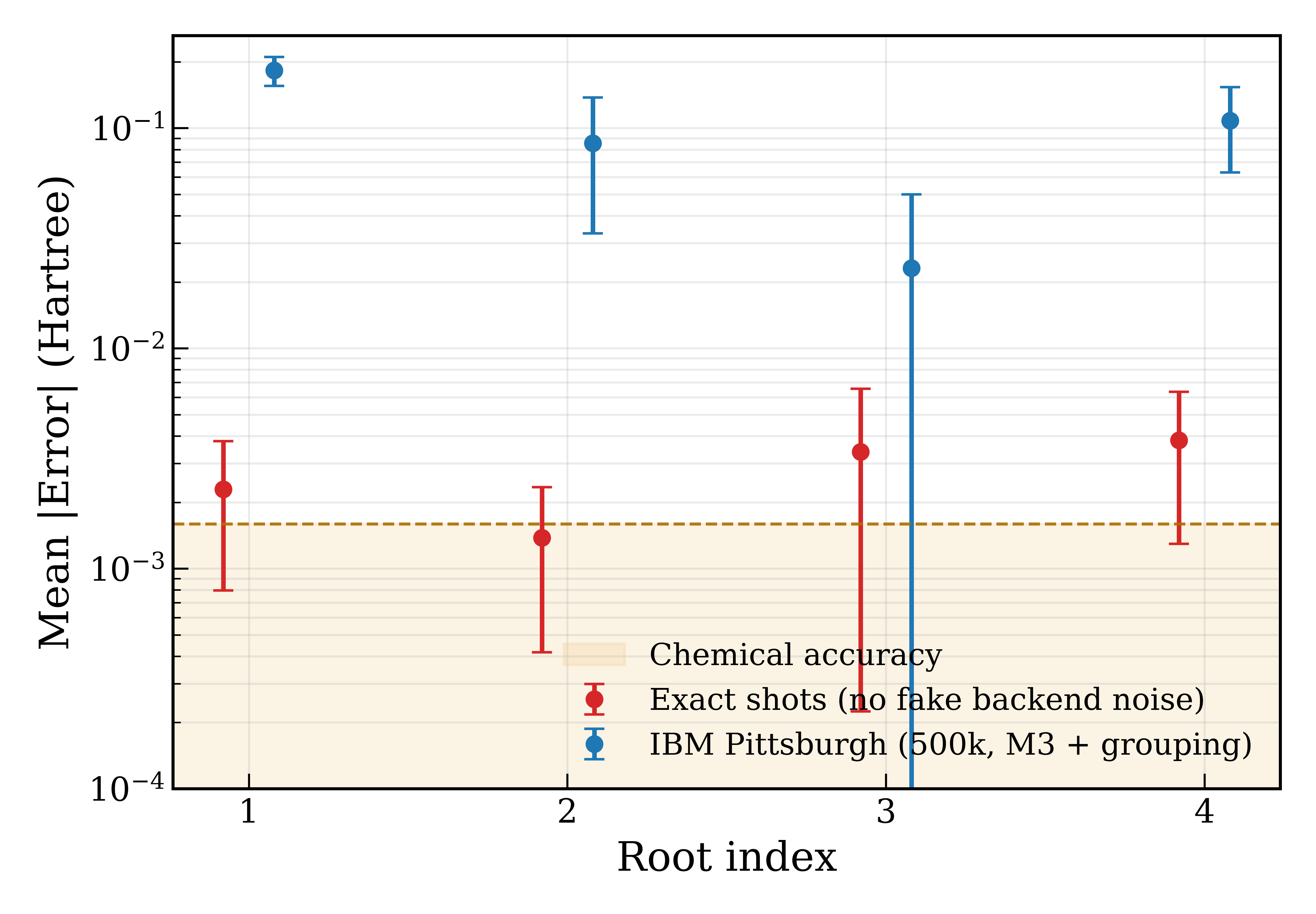}
    \caption{IBM Pittsburgh device}
    \end{subfigure}
\caption{Mean absolute energy error (Hartree) relative to FCI for the four roots obtained from q-sc-EOM for H$_2$ in minimal basis. Points show mean $\pm$ standard deviation over 5 repeated runs. The shaded region denotes the chemical accuracy threshold ($1.6\times10^{-3}$ Ha). These results demonstrate that hardware-induced systematic errors, rather than finite sampling, dominate the accuracy limit.}
\label{fig:h2_backend_compare}
\end{figure*}

\begin{figure}[t]
    \centering
    \includegraphics[width=\linewidth]{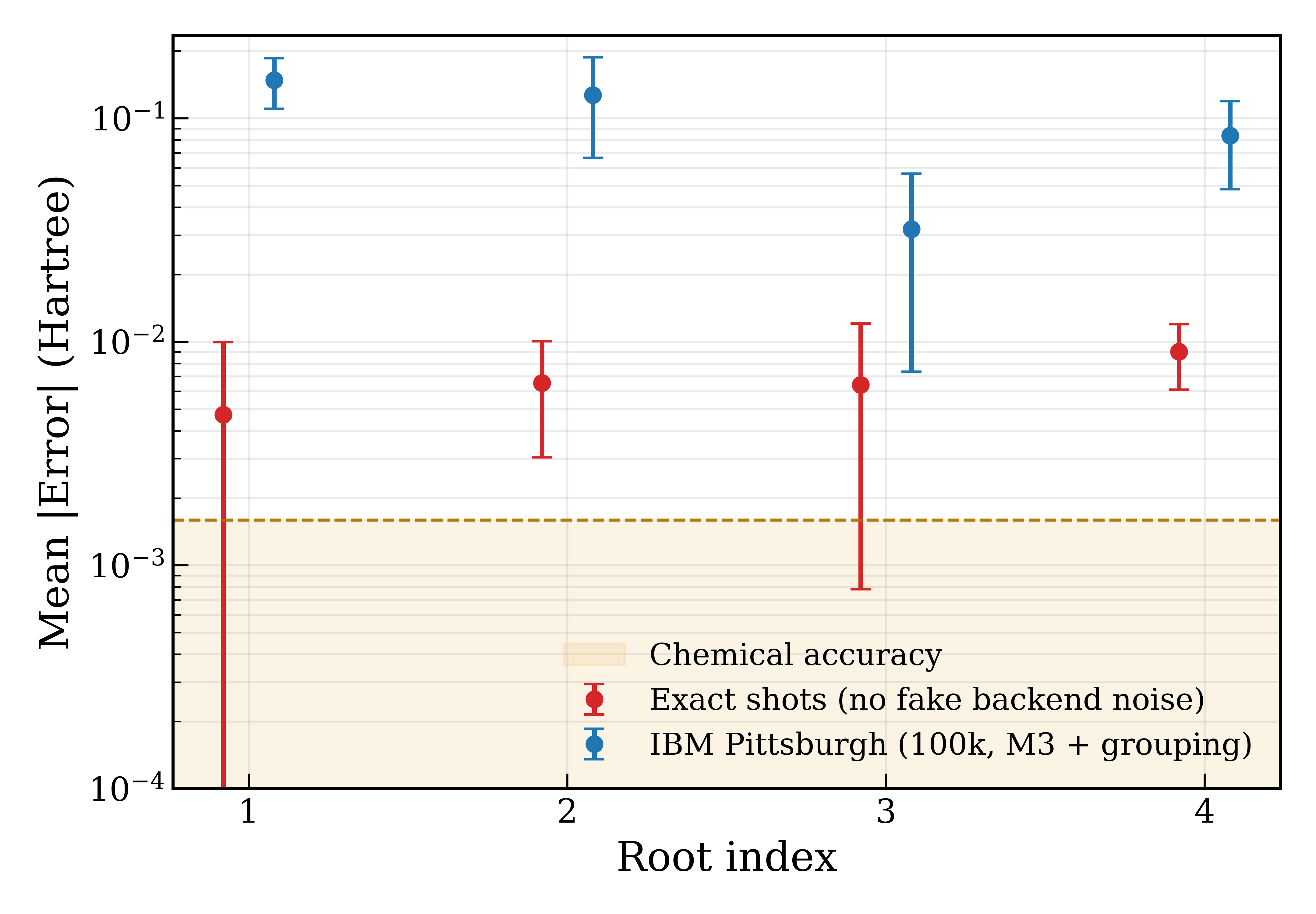}
\caption{Mean absolute energy error (Hartree) relative to FCI for the four roots obtained from q-sc-EOM for H$_2$ in minimal basis. Points show mean ± standard deviation over 5 repeated runs. The shaded region denotes the chemical accuracy threshold ($1.6\times10^{-3}$ Ha). These results demonstrate that errors due to shot noise are not the primary source of errors, as they do not affect the final computed values and may need sophisticated error mitigation techniques to mitigate gate errors.}
\label{fig:100shots}
\end{figure}
\begin{figure*}[t]
    \centering
    \begin{subfigure}[b]{0.45\textwidth}
    \includegraphics[width=\linewidth]{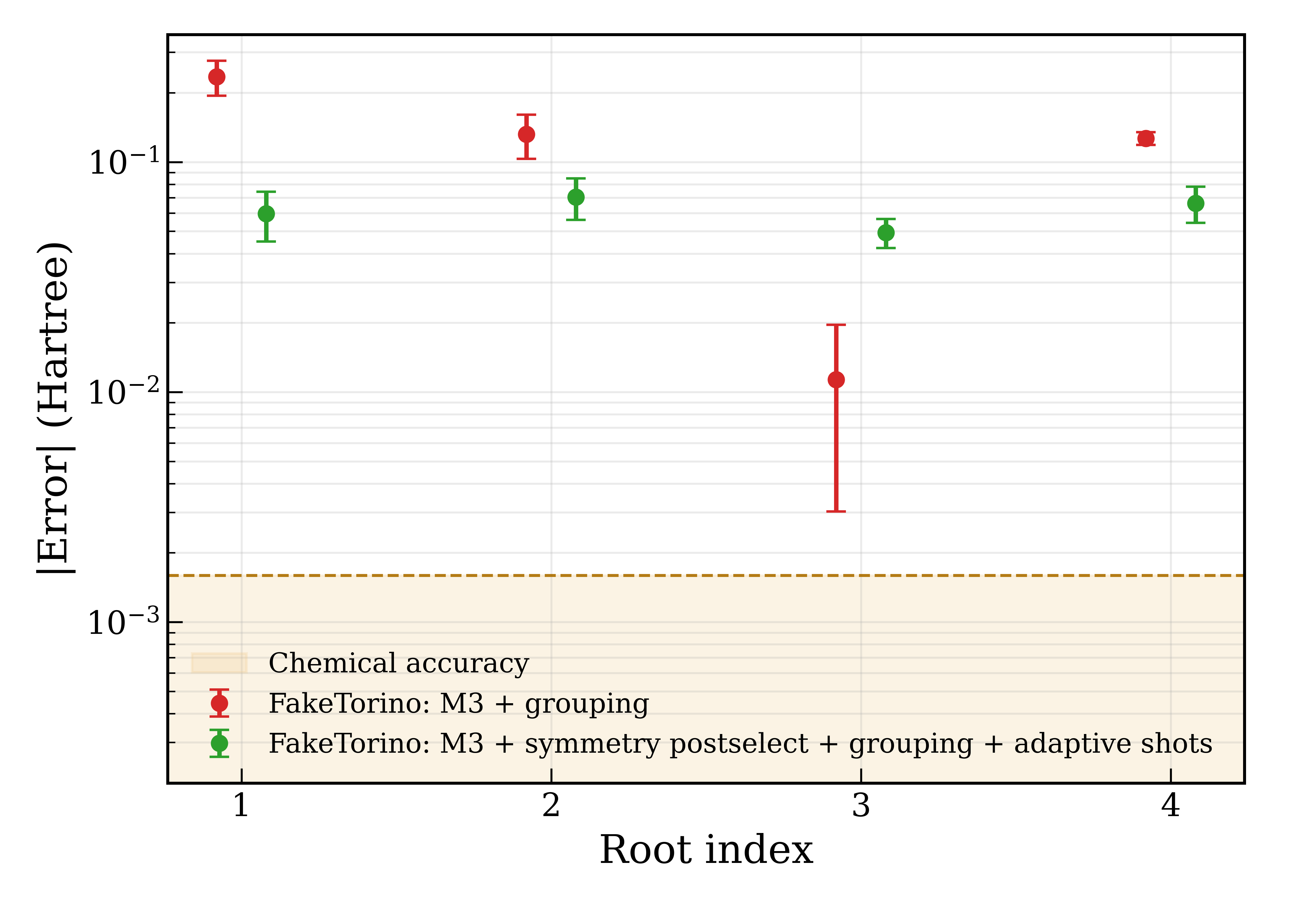}
    \caption{Fake Torino device }
        \end{subfigure}
    \begin{subfigure}[b]{0.45\textwidth}
    \includegraphics[width=\linewidth]{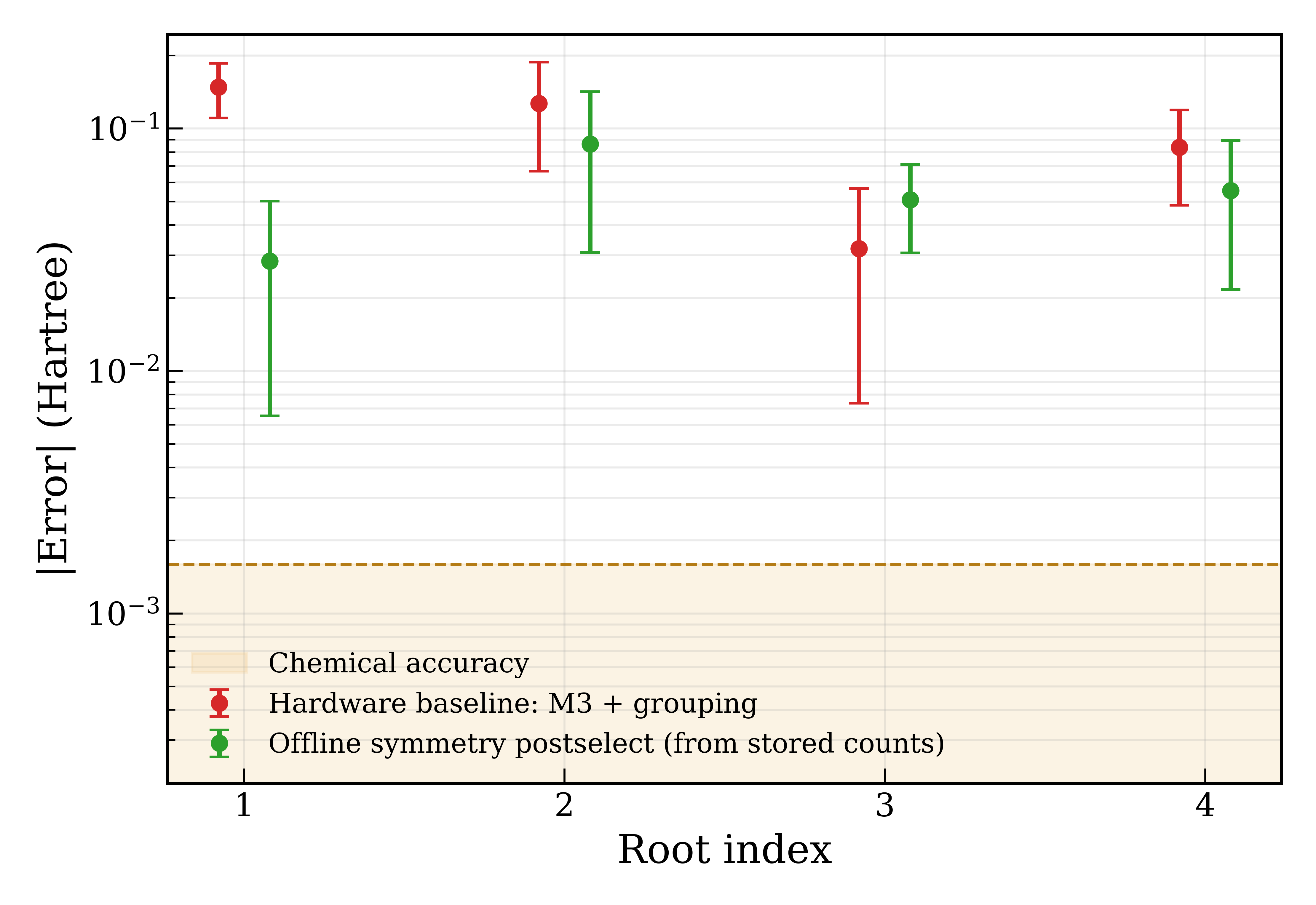}
    \caption{IBM Pittsburgh device}
    \end{subfigure}
  \caption{Root-wise q-sc-EOM error comparison on H$_2$ (STO-3G, \(R=0.74~\text{\AA}\)): baseline (M3 + grouping) vs corrected (M3 + symmetry postselection + grouping + adaptive shots). Error bars denote mean \(\pm\) one standard deviation over 5 runs at fixed grouped-shot budget (\(500\,000\) per run). (\(100\,000\) shots per run also yield quantitatively similar results, indicating saturation due to sampling errors.)}
  \label{fig:h2_sym_adaptive_fig8}
\end{figure*}

\subsubsection{Final Computational cost}

The computational cost of the ADAPT-VQE/LUCJ+q-sc-EOM scheme can be estimated separately for the ground state and excited state steps, which would be in addition to an O(N$^5$) shot scaling of ADAPT-VQE or O(N$^4$) shot scaling of LUCJ ansatz. The shot count estimate of the brute force implementation of the q-sc-EOM algorithm is O(N$^{12}$), while if the Davidson algorithm ~\cite{kim2023two} is used, the shot count scaling comes down to O(N$^8$) without any approximation. Further, further reduction using the BRG technique can bring it down to O(N$^5$). The circuit length of ADAPT-VQE depends on the number of iterations of ADAPT and can be studied further in the Ref. ~\cite{grimsley2019adaptvqe}, and the brute force implementation of q-sc-EOM adds a maximum of seven CNOT gates in the circuit to be implemented on top of the ground-state VQE circuit, which is a negligible addition compared with the ground-state circuit. The Davidson algorithm adds significant CNOT gates, where addition scales as O(N$^5$) to O(N$^6$), while a small increase using the BRG technique.

\subsection{Quantum hardware tests and error mitigation strategies}\label{rd3}

We present the mitigation stack used in q-sc-EOM benchmarks at different total shot budgets and the hardware experiments below. We present 5 statistically independent repeats per budget on each fake backend (FakeTorino and FakeBoston) and hardware.
Given q-sc-EOM basis states \(\{\ket{\psi_i}\}\), we estimate
\begin{equation}
M_{ij}=\bra{\psi_i}\hat{H}\ket{\psi_j},
\end{equation}
where diagonal elements are measured directly, and off-diagonal elements are reconstructed from phase-superposition circuits.
After estimation, \(M\) is enforced Hermitian via \(M\leftarrow(M+M^\dagger)/2\) before diagonalization. Error mitigation in each of these matrix elements is a problem similar to mitigating error in the ground-state circuit.

\subsubsection{Error-Mitigation Protocols}
\label{sec:h2_qsceom_mitigation_protocol}
We will make use of the following concepts in this study: 
\begin{enumerate}
\item \textbf{Pauli-basis grouping (measurement compression):}
For each q-sc-EOM setting, Pauli terms are grouped by common measurement basis (same non-identity Pauli axis pattern), so one circuit execution contributes to multiple Hamiltonian terms.
\item \textbf{M3 readout mitigation:}
Counts are corrected using matrix-free measurement mitigation (M3) based on local assignment calibrations.
For qubit \(q\), the local assignment model is
\begin{equation}
A_q=
\begin{pmatrix}
1-\epsilon_q & \epsilon_q\\
\epsilon_q & 1-\epsilon_q
\end{pmatrix},
\end{equation}
and corrected quasi-probabilities are used to compute Pauli expectations~\cite{nation2021scalable}.

\item \textbf{Symmetry postselection:}
Before expectation extraction, bitstrings are filtered to the target sector
$N_\alpha=1,\qquad N_\beta=1
\quad (\text{equivalently }N=2,\;S_z=0\text{ for H}_2)$.
In this implementation, postselection is applied to diagonal-in-\(Z\) (\(I/Z\)-only) measurement terms where particle-count constraints are directly evaluable from bitstrings~\cite{bonet2018low,sagastizabal2019experimental}.

\item \textbf{Adaptive shot allocation over q-sc-EOM:} Let \(s=(i,j,\phi)\) index each q-sc-EOM setting.
A pilot pass estimates a variance proxy
\begin{equation}
v_s\approx\sum_{\ell}|c_\ell|^2\left(1-\langle P_\ell\rangle_s^2\right),
\end{equation}
then the remaining shots are allocated approximately as
\begin{equation}
n_s\propto\sqrt{v_s},
\end{equation}
subject to a minimum per-setting floor and fixed total budget.
This concentrates shots on higher-variance settings (typically off-diagonal matrix elements)~\cite{arrasmith2020operator,gu2021adaptive}.

\end{enumerate}

%\subsubsection*{Raw vs Corrected definitions}
%\begin{itemize}
%\item \textbf{Raw:} Pauli grouping + M3 only, with legacy off-diagonal estimator, no symmetry postselection, and uniform shot allocation.
%\item \textbf{Corrected:} Pauli grouping + M3 + balanced 4-phase estimator + symmetry postselection + adaptive shot allocation.
%\end{itemize}

To keep fair comparisons across budgets, the target total shot budget $B$ is fixed. If $n_m$ is the number of matrix elements in the Hamiltonian matrix, $u$ is the number of grouped terms in the Hamiltonian,  and \(n\) is the shots allocated to each unit, the effective total shots are 
\begin{equation}
B_{actual}=u\times n \times n_m.
\end{equation}
%\subsubsection{Scope of mitigation in this study}
%The present protocol isolates readout/postprocessing and estimator-level mitigation.
%It does not include gate-noise-focused methods such as zero-noise extrapolation, probabilistic error cancellation, or dynamical decoupling; those are deferred to separate hardware-focused studies.

\subsubsection {Hardware Results}

Figure~\ref{fig:h2o_hw_roots} shows a hardware demonstration of q-sc-EOM for H$_2$O in a reduced active space (2e,2o) at a fixed budget of 100,000 shots per run on IBM Pittsburgh (5 independent runs) and LUCJ ansatz for the ground-state circuit. We have used IBM Pittsburgh hardware for this study, specifically the qubit numbers 69, 70, 71, 72, which were the best performing at the time of our study. These machine and qubit IDs are constant throughout our study. We may use different fake backends, but the same hardware. This is because, in our limited access to IBM at the time of running these experiments, IBM Pittsburgh gave us the best performance, equally good compared with some other devices, and was more readily available. The left panel reports root-wise absolute error, and the right panel reports the corresponding energies, enabling direct separation of statistical spread and systematic bias across states. Under the baseline mitigation strategy (Pauli grouping + M3), we observe state-dependent performance: some roots remain comparatively well behaved while others retain larger deviations from the exact reference. This root selectivity is consistent with the subspace-diagonalization structure of q-sc-EOM, where non-uniform matrix-element noise propagates differently to different eigenvalues. If we focus only on the excited states, roots 2-4, they have an error in the order of $5\times 10^{-2}$ Ha or 50 mHa, which is above ideal for chemical usefulness. This experiment exposes the importance of mitigating gate errors to reach useful numbers using present hardware.

Figure~\ref{fig:h2_backend_compare} shows that for H$_2$ in the STO-3G basis, shot noise alone does not prevent q-sc-EOM from reaching near-chemical accuracy at 500k shots. The exact-shot data display errors consistent with statistical sampling fluctuations. Introducing realistic noise through the FakeTorino backend increases the error significantly, which is not mitigated much using readout error mitigation by M3, indicating that stochastic noise and readout errors (partially mitigated by M3) are not the principal limiting factors at this circuit scale, but rather the gate errors are the really challenging part. Execution on IBM Pittsburgh hardware leads to errors that are very similar. The circuit for the ground state in this study has a total of 305 gates with 65 CNOT gates and 192 Rz gates. Although this is not a very deep circuit, the gate errors here still have a significantly large impact on the errors. The persistence of a systematic bias despite high shot counts indicates that coherent gate errors, crosstalk, leakage, and time-dependent calibration drift dominate. Because q-sc-EOM relies on subspace diagonalization, small systematic biases in Hamiltonian matrix elements can be amplified through the eigenvalue problem, leading to disproportionately larger energy deviations. Thus, even for minimal-basis H$_2$, current hardware performance is limited primarily by structured device noise rather than finite sampling statistics.

In Figure~\ref{fig:100shots}, we evaluate whether part of the error in the fakebackend and hardware is arising because of sampling error. A Pittsburgh hardware run with 100,000 total shots per run confirms that the sampling errors are almost not contributing at all. The plot has red dots and bars that represent a simulation with only sampling errors, which is significantly above chemical accuracy compared with the Fig. \ref{fig:h2_backend_compare}, but still the blue points are similar in errors as Fig. \ref{fig:h2_backend_compare} confirming that the cause is truly errors at the measurement or the gate level and not at sampling.

For more advanced error gate level error mitigation, we tried many strategies including the use of Twirling (see \ref{a2}) and Zero Noise extrapolation, but none of the strategies had any measurable effects in our case, possibly because of a large number of CNOT gates, but the best results we were able to achieve was with post measurement symmetry projection. Figure~\ref{fig:h2_sym_adaptive_fig8} present root-wise q-sc-EOM errors with only readout error mitigation and readout along with symmetry projection-based error mitigation. We carried out this experiment using both the FakeTorino device (left panel) and IBM Pittsburgh hardware (right panel). These results are for H$_2$ (STO-3G, \(R=0.74~\text{\AA}\)). Reported points are means with one-standard-deviation error bars over 5 repeated runs at fixed grouped-shot budget \(500\,000\) shots per run.
The key observation here is not a uniform downward shift of all roots, but a root-dependent redistribution of error. Several roots improve markedly after correction, while at least one root can remain unchanged or worsen slightly, depending on backend/noise realization. This behavior aligns with our understanding of subspace excited-state methods, where mitigation changes the statistical and systematic error balance of individual matrix elements entering the projected Hamiltonian and overlap problems, resulting in different effects on individual states. However, overall projection-based error mitigation brings the overall errors of each state below 100 mHa and closer to the order of 50mHa. 

\section{Conclusion} \label{Conclusion}
In this work, we developed and tested a practical quantum workflow for molecular excited states by combining high-quality variational ground-state preparation with the ADAPT-VQE and LUCJ ansätze and excited states using the q-sc-EOM subspace method. We applied it to challenging two-bond-breaking potential energy surfaces in NH$_3$ and H$_2$O molecules. We compared the performance of these methods on a perfect quantum device against exact results in the active space and the scalable classical EOM-CCSD approach, while also studying the real resource bottlenecks of the excited-state calculations. To make the method more realistic for larger systems, we evaluated the incorporation of Davidson subspace diagonalization and basis-rotation grouping, and further examined hardware execution with readout mitigation and gate-error mitigation strategies.  

We found that ADAPT-VQE + q-sc-EOM can produce accurate excited-state surfaces in bond-breaking regimes where the classical EOM-CCSD description deteriorates, because the quantum reference captures the multireference character of the ground state more faithfully. At the same time, the combination of Davidson and basis-rotation grouping reduces the effective measurement scaling of the excited-state stage from O(N$^{12}$) to O(N$^{5}$), showing that the method can be made far more resource-efficient without sacrificing accuracy. On hardware, M3 and symmetry postselection improved results, but the dominant factor limiting accuracy was identified as gate-level device noise, with typical hardware excited-state deviations on the order of 50 mHa in the demonstrated settings, with the symmetry-projection technique reducing errors.  

Taken together, these results show that quantum subspace methods for excited states are not only theoretically appealing but are beginning to look algorithmically viable for chemically relevant problems. The present study points to a clear path forward for how to reach usefully accurate excited states while reducing quantum resources. At the same time, pushing on the accurate hardware implementation, which may require substantially better gate fidelities and more effective error mitigation techniques for excited states to reach accuracies that are useful for computational studies. More broadly, this work strengthens the case that quantum computing may become especially valuable in the regimes that challenge conventional excited-state methods most—multi-reference ground states, bond-breaking regimes, and excited states dominated by higher excitations —thereby advancing the long-term goal of predictive quantum chemistry for molecules, materials, and photochemical processes.

\section{Data availability}
The data in this paper are produced using the codes available in the open source repository at https://github.com/asthanalab/QCANT

\section{Acknowledgment}
 AA and SPS acknowledge funding from NSF award numbers 2427046 and 2429752, the UND startup funds and funding from Moderna Inc. All authors acknowledge the IBM Quantum grant and the UND Computational Research Center for computing resources.

 \appendix
 
 \section{Choice of ansatz in ground state VQE}
 We explore the capabilities of various ansätze and compare them with the Full configuration interaction (FCI) energy values. As a challenging test case for correlation energy, we use the linear $\mathrm{H_4}$ system and compute its energy using multiple ansatz choices, including UCCSD-VQE, HEA, and ADAPT-VQE with the STO-6G basis set. In this scenario, $\mathrm{H_4}$ has 4 spatial orbitals and 8 qubits, and the initial state is the Hartree--Fock state for chemically inspired ansatz choices (UCCSD and ADAPT-VQE).  For HEA, we utilized 4 layers of rotation (Ry) and cascade connectivity of CNOT gates to generate the ansatz.

\begin{figure}[thp]
     \centering
     \includegraphics[width=\linewidth]{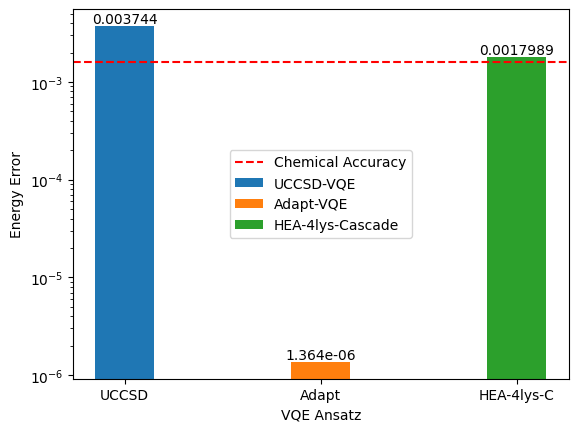}
     \caption{Comparison of errors $E_{\mathrm{ansatz}} - E_{\mathrm{FCI}}$ for various ansatz (UCCSD, ADAPT and HEA) for $\mathrm{H_4}$ linear-3A (strongly correlated system). The dashed red line represents "chemical accuracy" as 0.00159 Ha (1 kcal/mol).}
     \label{1}
 \end{figure}

 The energy difference for each ansatz was calculated with respect to FCI and is represented in Fig. \ref{1}.  The horizontal line represents the chemical accuracy. Both the UCCSD and the HEA fail to achieve errors below chemical accuracy. The dynamic ansatz (ADAPT-VQE) works better than the other ansatz studied in the present conditions. ADAPT-VQE uses fewer parameters in comparison to UCCSD to achieve better accuracy. Our results are in agreement with the literature~\cite{anand2022quantum,mullinax2024classical,PRXQuantum.2.020310}, proving that ADAPT-VQE surpasses UCCSD in the treatment of strongly correlated systems.

 VQEs are optimizer-dependent in nature. They can be classified into gradient and non-gradient based optimizers. For this study, we tested our linear $\mathrm{H_4}$ system with L-BFGS (gradient) and COBYLA, Powell (non-gradient) optimizers. In the exact scenario (infinite shot simulation), we found that all the tested optimizers achieved similar accuracy levels.
 %Since this work involves the implementation of an algorithm  quantum computer, testing these optimizers in noisy environments is of utmost relevance.
 We simulated our $\mathrm{H_4}$-Linear system in the presence of shot noise.  The chosen gradient based optimizer (L-BFGS) failed in a noisy scenario, these results are in agreement with recent literature studies ~\cite{motta2024quantum,lavrijsen2020classical}. The results showed both non-gradient optimizers Powell and COBYLA were able to achieve really low root-mean-square deviation (RMSD) compared to exact scenarios.  On the basis of these results, we decided to utilize ADAPT-VQE with Powell as an optimizer to generate the unitary operator required for excited states.

 \begin{figure}[htp]
    \centering
    \includegraphics[width=\linewidth]{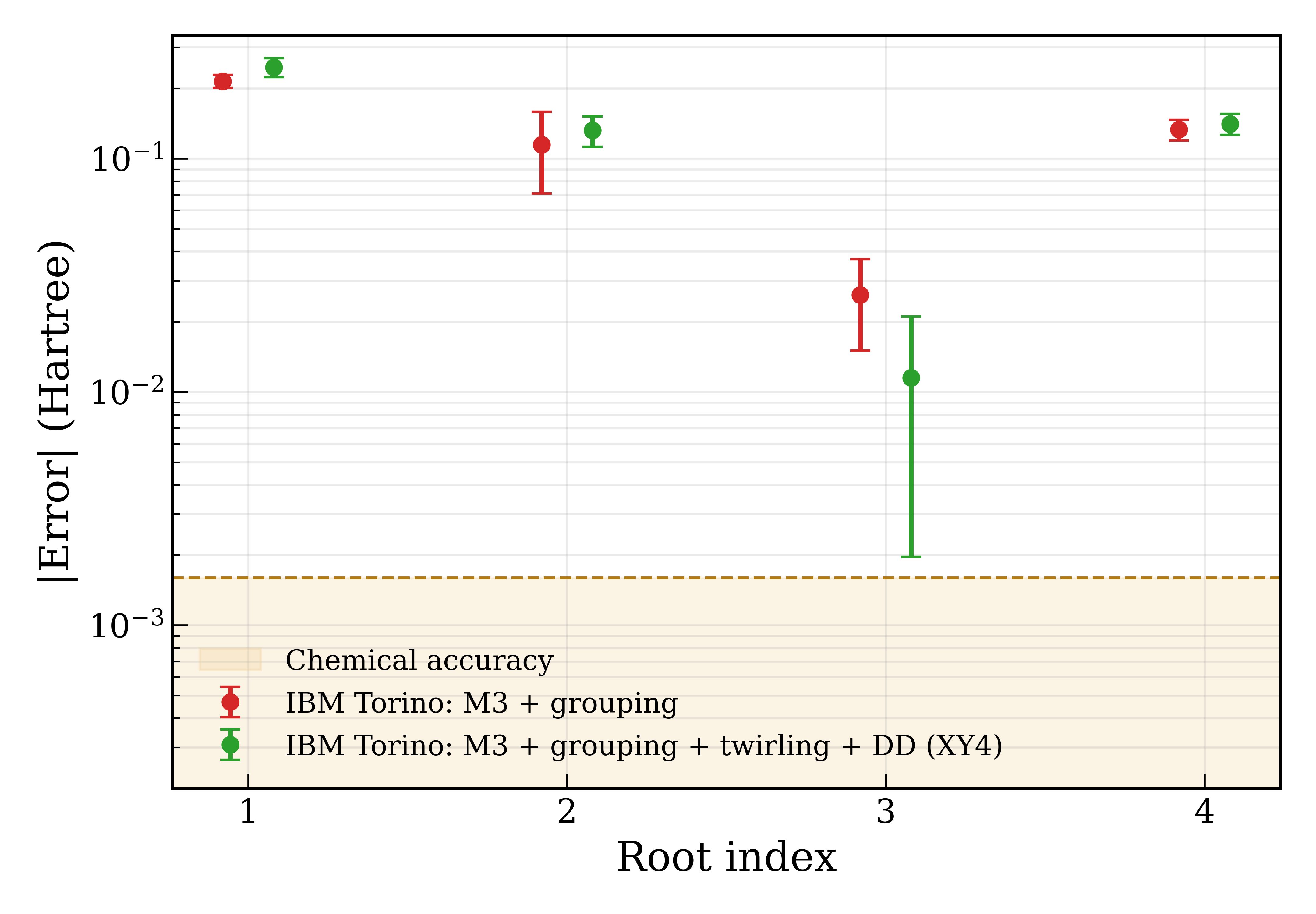}
\caption{Mean absolute energy error (Hartree) relative to FCI for the four roots obtained from q-sc-EOM for H$_2$ in minimal basis on IBM Torino using twirling +dynamic decoupling error mitigation. Points show mean $\pm$ standard deviation over 5 repeated runs. The shaded region denotes the chemical accuracy threshold ($1.6\times10^{-3}$ Ha). These results demonstrate that dynamic decoupling is unable to tackle the gate-level errors present in this calculation.}
\label{fig:twirling}
\end{figure}

 \section{Twirling, DD(XY4)}\label{a2}
 We benchmarked runtime-level gate-noise mitigation on hardware (in Fig. \ref{fig:twirling}) for H$_2$ (STO-3G, $R=0.74\,\text{\AA}$) using q-sc-EOM with Pauli grouping and M3 readout mitigation as the baseline, and compared against a mitigated configuration with gate twirling plus dynamical decoupling (DD, XY4). Figure~\ref{fig:twirling} summarizes 5 independent runs at a fixed budget of 100,000 shots per run on IBM Torino. The mitigation effect is root-dependent: one state shows strong improvement, while others degrade modestly, yielding no net gain in aggregate mean error. This behavior is consistent with mitigation reshaping coherent/stochastic error channels differently across observables, rather than uniformly lowering all matrix-element errors. The result highlights that for q-sc-EOM excited-state spectra, twirling+DD should be treated as a tunable protocol (sequence, randomization level, and scheduling) and validated per root family and shot regime.

\bibliography{main}% Produces the bibliography via BibTeX.
\end{document}